\pdfoutput=1

\documentclass[11pt,twoside,a4paper,cmspaper,final,collab]{cms-tdr}
\def\svnVersion{374588}\def\cmsCernNoTag{CERN-EP/2016-199}\def\cmsCernDate{\today}\def\cmsMessage{Published in the Journal of High Energy Physics as \href{http://dx.doi.org/10.1007/JHEP11(2016)055}{\doi{10.1007/JHEP11(2016)055.}}}

\begin{document}\cmsNoteHeader{HIN-15-011}

\hyphenation{had-ron-i-za-tion}
\hyphenation{cal-or-i-me-ter}
\hyphenation{de-vices}
\RCS$Revision: 367073 $
\RCS$HeadURL: svn+ssh://svn.cern.ch/reps/tdr2/papers/HIN-15-011/trunk/HIN-15-011.tex $
\RCS$Id: HIN-15-011.tex 367073 2016-09-08 16:51:09Z alverson $
\newlength\cmsFigWidth
\ifthenelse{\boolean{cms@external}}{\setlength\cmsFigWidth{0.85\columnwidth}}{\setlength\cmsFigWidth{0.4\textwidth}}
\ifthenelse{\boolean{cms@external}}{\providecommand{\cmsLeft}{top\xspace}}{\providecommand{\cmsLeft}{left\xspace}}
\ifthenelse{\boolean{cms@external}}{\providecommand{\cmsRight}{bottom\xspace}}{\providecommand{\cmsRight}{right\xspace}}
\providecommand{\pttrk}{\ensuremath{\pt^\text{trk}}\xspace}
\providecommand{\HYDJET}{\textsc{hydjet}\xspace}
\cmsNoteHeader{HIN-15-011}
\title{Decomposing transverse momentum balance contributions for quenched jets in PbPb collisions at $\sqrt{s_\mathrm{NN}} = 2.76$\TeV}

\date{\today}
\abstract{

Interactions between jets and the quark-gluon plasma produced in heavy ion collisions are studied via the angular distributions of summed charged-particle transverse momenta (\pt) with respect to both the leading and subleading jet axes in high-\pt dijet events.  The contributions of charged particles in different momentum ranges to the overall event \pt balance are decomposed into short-range jet peaks and a long-range azimuthal asymmetry in charged-particle \pt.  The results for PbPb collisions are compared to those in pp collisions using data collected in 2011 and 2013, at collision energy $\sqrt{s_\mathrm{NN}} = 2.76$\TeV with integrated luminosities of 166\mubinv and 5.3\pbinv, respectively, by the CMS experiment at the LHC.  Measurements are presented as functions of PbPb collision centrality, charged-particle \pt, relative azimuth, and radial distance from the jet axis for balanced and unbalanced dijets.

}

\hypersetup{%
pdfauthor={CMS Collaboration},%
pdftitle={Decomposing transverse momentum balance contributions for quenched jets in PbPb collisions at sqrt(s[NN]) = 2.76 TeV},%
pdfsubject={CMS},%
pdfkeywords={CMS, physics, heavy ions, jet quenching}%
}

\maketitle
\clearpage

\hyphenation{di-a-me-ter re-con-struc-tion using}

\section{Introduction}

High transverse momentum (\pt) jets originating from partons produced in the initial hard scatterings in ultra-relativistic heavy ion collisions have been used successfully to study the properties of the quark-gluon plasma (QGP)~\cite{Bjorken:1982tu}.  The observation of the jet quenching phenomenon, first at the BNL RHIC~\cite{Adams:2006yt, Adare:2007vu} and then at the CERN LHC~\cite{Aad:2010bu, Chatrchyan:2011sx, Chatrchyan:2012nia, Adam:2015ewa}, began the era of detailed experimental studies trying to assess both the redistribution of energy from the parton as it interacts with the QGP, and the possible QGP response to the propagating parton.   In these studies, a suppression of strongly interacting hard probes has been observed, including suppression of charged-particle yields associated with jets when compared with pp data at the same center-of-mass energy.  The dependence of jet quenching on the collision centrality (i.e. the degree of the overlap of the two colliding nuclei, with fully-overlapping nuclei defined as ``0\% central''), has also been established, with stronger quenching effects reported for more central collisions.  Detailed studies at CMS and ATLAS report that not only the suppression of jet yields, but also the relative differences between the momenta of the leading and subleading jets in dijet events increase in more central collisions, indicating that the jet quenching effect depends on the path length of the parton traversing the medium~\cite{Aad:2010bu, Chatrchyan:2011sx, Chatrchyan:2012nia}. Corresponding measurements from the most peripheral (i.e. least central) PbPb data in these studies and from pPb collisions~\cite{Chatrchyan:2014hqa} are found to be similar to those in pp collisions,  implying that the jet quenching phenomenon is caused by hot nuclear matter effects.   Recently, CMS reported results showing the difference in the distribution of charged-particle \pt between the subleading and leading jet hemispheres in PbPb events with asymmetric dijets~\cite{mpt}.  In these results, it is found that low-\pt particles extending to large angles away from the dijet axis in the subleading hemisphere must be considered in order to recover the momentum balance in these events. In addition to these momentum balance studies, precise measurements of the fragmentation pattern~\cite{CMS:2012vba} and the distribution of charged-particle \pt as a function of radial distance from the jet axis~\cite{Chatrchyan:2013kwa}, have also shown that the jet structure is modified by the medium.  These modifications extend to large distances in relative pseudorapidity ($\Delta\eta$) and relative azimuth ($\Delta\phi$) with respect to the jet axis~\cite{HIN_2014_016}.  These studies have found softening of the jet fragmentation in PbPb collisions with respect to pp events, with the most significant excess of soft-hadron yields observed in more central PbPb events.

The analysis presented in this paper probes the details of the momentum distribution in dijet events, taking advantage of the high production rates for dijet events at the LHC, and the CMS detector's ability to measure charged-particle tracks over an extended $\eta$ and \pt range.  Two-dimensional correlations between the reconstructed jets and charged-particle tracks (jet-track correlations) are constructed in $\Delta\eta$ and $\Delta\phi$.  These correlations are used to decompose the overall event \pt distribution into three components:  two 2D Gaussian-like peaks associated with the leading and subleading jets, and an azimuthal asymmetry in the distribution of momentum under the jet peaks.  Results for each component are presented as a function of $\Delta\phi$, and the jet momentum density profile (``jet shape'') is also presented as a function of the radial distance from the jet axis in the $\Delta\eta$-$\Delta\phi$ plane.  Measurements are performed differentially in collision centrality, charged-particle transverse momentum (\pttrk), and dijet asymmetry.  These detailed differential studies provide input for theoretical models that attempt to describe the patterns of energy loss by a highly-energetic probe passing through the QGP.

The data used in this analysis are from PbPb collisions at a nucleon-nucleon center-of-mass energy of 2.76\TeV, corresponding to an integrated luminosity of 166\mubinv.  For the reference measurement, pp data taken in 2013 at the same energy corresponding to an integrated luminosity of 5.3\pbinv are used. These studies allow for a detailed characterization of the two-dimensional (in $\Delta\eta$ and $\Delta\phi$) \pttrk distributions for charged-particle tracks with respect to the jet axes, providing information about the topology of the event from the jet perspective and details about the \pttrk flow modification in dijet events.

In this paper, Section 2 gives general information about the CMS detector, and Section 3 outlines jet and track reconstruction procedures for PbPb and pp data.  Section 4 describes the selection of events, while Section 5 details the procedure applied to analyze these events and evaluate systematic uncertainties.  Section 6 presents results as a function of $\Delta r$ (Section 6.1), $\Delta \phi$ (Section 6.2), and integrated transverse momentum balance over the whole event (Section 6.3).  Finally, Section 7 summarizes and concludes the paper.

\section{The CMS detector}
\label {sec:Detector}

The central feature of the CMS apparatus is a superconducting solenoid of 6\unit{m} internal diameter, providing a magnetic field of 3.8\unit{T}. Within the solenoid volume are a silicon pixel and strip tracker, a lead tungstate crystal electromagnetic calorimeter (ECAL), and a brass and scintillator hadron calorimeter (HCAL), each composed of a barrel and two endcap sections.  Two hadronic forward (HF) steel and quartz-fiber calorimeters complement the barrel and endcap detectors, providing coverage up to $\abs{\eta}<5.2$. In this analysis, the collision centrality is determined using the total sum of transverse energy (\ET) from calorimeter towers in the HF region (covering \mbox{$2.9<\abs{\eta}<5.2$}). The \ET distribution is used to divide the event sample into bins, each representing 0.5\% of the total nucleus-nucleus hadronic interaction cross section.  A detailed description of centrality determination can be found in Ref.~\cite{Chatrchyan:2012nia}.

Jet reconstruction for this analysis relies on calorimeter information from the ECAL and HCAL.  For the central region ($\abs{ \eta } < 1.74$) from which jets are selected for this analysis, the HCAL cells have widths of 0.087 in both $\eta$ and $\phi$. In the $\eta$--$\phi$ plane, and for $\abs{\eta} < 1.48$, the HCAL cells map on to $5{\times}5$ ECAL crystal arrays to form calorimeter towers projecting radially outwards from close to the nominal interaction point. Within each tower, the energy deposits in ECAL and HCAL cells are summed to define the calorimeter tower energies, subsequently used to provide the energies and directions of hadronic jets~\cite{Chatrchyan:2011ds}.

Accurate particle tracking is critical for measurements of charged-hadron yields.  The CMS silicon tracker measures charged particles within the range $\abs{\eta}< 2.5$. It consists of 1,440 silicon pixel and 15,148 silicon strip detector modules. For nonisolated particles of $1 < \pt < 10\GeV$ and $\abs{\eta} < 1.4$, the track resolutions are typically 1.5\% in \pt and 25--90 (45--150)\mum in the transverse (longitudinal) impact parameter \cite{TRK-11-001}.  Performance of the track reconstruction in pp and PbPb collisions will be discussed in Section~\ref{sec:Jetreco}.

A detailed description of the CMS detector, together with a definition of the coordinate system used and the relevant kinematic variables, can be found in ~\cite{Chatrchyan:2008zzk}.

\section{Jet and track reconstruction}
\label{sec:Jetreco}

For both pp and PbPb collisions, jet reconstruction in CMS is performed with the anti-\kt algorithm, as implemented in the FastJet framework \cite{Cacciari:fastjet1,bib_antikt}, using a distance parameter $R=0.3$.  Jets are reconstructed offline (i.e. after raw data is recorded) based on energy deposits in the CMS calorimeters. Raw jet energies are obtained from the sum of the tower energies and raw jet momenta from the vectorial sum of the tower momenta, and are corrected to establish a relative uniform response of the calorimeter in $\eta$ and a calibrated absolute response in \pt. For PbPb collisions, the CMS algorithm ``HF/Voronoi'' is used to estimate and subtract the heavy-ion underlying event based on information from HF energy measurements as well as Voronoi decomposition of particle flow~\cite{CMS-DP-2013-018, mpt}.  For pp collisions, the contribution from the underlying event is negligible and no underlying event subtraction is employed.

Monte Carlo (MC) event generators have been used for evaluation of the jet and track reconstruction performance. Jet events are generated by the
\PYTHIA MC generator~\cite{bib_pythia} (version 6.423, tune
    Z2~\cite{Field}). Simulated events are further propagated through the
CMS detector using the \GEANTfour package~\cite{GEANT4} to simulate
the detector response. In order to account for the influence of the underlying PbPb event, the \PYTHIA events are
embedded into fully simulated PbPb events, generated by \HYDJET~\cite{Lokhtin:2005px} (version 1.8) that is tuned to reproduce the total particle multiplicities, charged-hadron spectra, and elliptic flow at all centralities. The embedding is done by mixing the simulated signal information from \PYTHIA and \HYDJET, hereafter referred to as \PYTHIA{}+\HYDJET. These events are then propagated through the same jet and track reconstruction and analysis procedures as pp and PbPb data.

\sloppypar{The jet energy scale (JES) is established for pp using \PYTHIA events and for PbPb using  \PYTHIA{}+\HYDJET events in classes of event centrality.  The accuracy of the reconstruction and correction procedure is tested as a function of jet \pt and $\eta$, by comparing a sample of reconstructed and corrected jets to the jets originally simulated in that sample.  To account for the dependence of the JES on the fragmentation of jets, an additional correction is applied as a function of reconstructed jet \pt and as a function of the number of tracks with $\pt > 2$\GeV within a radius $\Delta r < 0.3$ around the jet axis. This correction is derived separately for pp and PbPb data, as described in Ref.~\cite{mpt}.}

For studies of pp data and \PYTHIA simulation, tracks are reconstructed using the same iterative method~\cite{TRK-11-001} as in the previous CMS analyses of pp collisions. For PbPb data and \PYTHIA{}+\HYDJET simulation, a dedicated heavy-ion iterative track reconstruction method~\cite{CMS:2012aa,Chatrchyan:2013kwa} is employed.  Tracking efficiency for charged particles in pp collisions ranges from approximately 80\% at $\pt\approx0.5$\GeV to 90\% or better at $\pt\approx10$\GeV and higher.  Track reconstruction is more difficult in the heavy-ion environment due to the high track multiplicity, and tracking efficiency for PbPb collisions ranges from approximately 30\% at 0.5\GeV to about 70\% at 10\GeV.  Detailed studies of tracking efficiency and of tracking efficiency corrections (derived as a function of centrality, \pttrk, $\eta$, $\phi$, and local charged particle density) can be found in Ref.~\cite{mpt}.

\section{Event selection}
\label{sec:evtsel}

The events for this analysis are selected using the CMS high-level trigger (HLT), with an inclusive single-jet trigger with a threshold of $\pt > 80$\GeV~\cite{HLT}.  This trigger is fully efficient in both PbPb and pp data for events containing offline reconstructed jets with $\pt > 120$\GeV. In order to suppress noncollision-related noise due to sources such as cosmic rays and beam backgrounds, the events used in this analysis are also required to satisfy offline selection criteria as documented in Refs.~\cite{Chatrchyan:2012nia,Chatrchyan:2012gt}. These include restricting PbPb events to those containing a reconstructed vertex with at least two tracks and a $z$ position within 15 cm of the detector center, and in which at least 3\GeV energy is deposited in at least three HF calorimeter towers on each side of the interaction point.

A dijet sample is selected using criteria matched to those of previous CMS analyses measuring dijet energy balance and correlated yields to high-\pt jets~\cite{Chatrchyan:2011sx,mpt, HIN_2014_016}.  In this selection, events are first required to contain a leading calorimeter jet with $p_\mathrm{T,1} > 120$\GeV in the range of $\abs{\eta_\text{jet}} < 2$, and a subleading jet of $p_\mathrm{T,2} > 50$\GeV, also in $\abs{\eta_\text{jet}}<2$.  Once the leading and subleading jets in the event have been identified, a tighter $\abs{\eta_\text{jet}}$ selection is applied to ensure stable jet reconstruction performance and good tracker acceptance for tracks on all sides of each jet:  only events in which both leading and subleading jets fall within $\abs{\eta_\text{jet}}<1.6$ are included in the final selected data sample.  The azimuthal angle between the leading and subleading jets is required to be at least $5\pi/6$. No explicit requirement is made either on the presence or absence of a third jet in the event. This jet sample is further divided based on the asymmetry between the leading and subleading jets, $A_\mathrm{J} = (p_\mathrm{T,1}-p_\mathrm{T,2})/(p_\mathrm{T,1}+p_\mathrm{T,2})$.  Two asymmetry classes are considered:  the more balanced part (defined by $A_\mathrm{J} < 0.22$) of the total dijet sample, and the more unbalanced part (defined by $A_\mathrm{J} > 0.22$).  The dividing value $A_\mathrm{J} = 0.22$ is chosen for consistency with previous CMS analyses~\cite{mpt, Chatrchyan:2011sx}.  In this analysis, 52\% of PbPb events are balanced, while 67\% of pp events are balanced.  For the PbPb data, the centrality of the collisions is also considered and results are compared for central events (with centrality 0--30\%) versus peripheral events (with centrality 50-100\%).

\section{Analysis procedure}

Dijet events in this analysis are studied differentially in collision centrality, with the following bins:  0--30\% (most central), 30--50\% (not shown), and 50--100\% (most peripheral). This analysis follows the procedure established in Refs.~\cite{HIN13002,HIN_2014_016}:  two-dimensional $\Delta \eta$--$\Delta \phi$ correlations with respect to the measured subleading and leading jet axes are constructed for charged-particle tracks in the event with $\pttrk \geq 0.5$\GeV and $\abs{\eta_\text{track}}<2.4$, in several \pttrk bins.  These correlations are weighted by \pttrk on a per-track basis, and normalized by the number of jets in the sample.  This produces two-dimensional $\Delta\eta$--$\Delta\phi$ average per-jet distributions of \pttrk with respect to the leading and subleading jets.
After the construction of the initial two-dimensional correlations described above, the remaining analysis procedure consists of the following steps, which will be discussed in detail below:
\begin{itemize}
\item A pair-acceptance correction, derived  by the ``mixed event'' method~\cite{HIN13002,HIN_2014_016};
\item The separation of correlations into jet-peak and long-range components;
\item Corrections for jet reconstruction biases:  a full simulation-based analysis is conducted to determine and subtract the correlated yield produced by jet selection bias.
\end{itemize}

\subsection{Pair-acceptance correction}
With jet acceptance of $\abs{\eta_\text{jet}}<1.6$, many tracks within $\abs{\Delta\eta}<2.5$ of a jet will fall outside of the track acceptance of $\abs{\eta_\text{track}}<2.4$, resulting in correlation geometry that falls with increasing $\Delta\eta$.  To correct for this pair-acceptance effect, a mixed-event distribution is constructed by correlating jets from the jet-triggered event sample with tracks from a sample of minimum bias events, matched in vertex position (within 1 cm) and collision centrality (within 2.5\%), following the technique used in Refs.~\cite{ppRidge,PbPbridge,pPbridge}.  In the following,  $N_\text{jets}$ denotes the number of dijet events selected as described in a given data sample. The per-jet associated yield, weighted per-track by \pttrk is defined as:

  \begin{equation}
  \label{2pcorr_incl}
  \frac{1}{N_\text{jets}}\frac{\rd^{2}\sum \pt}{\rd\Delta\eta\, \rd\Delta\phi}
  = \frac{ME(0,0)}{ME(\Delta\eta,\Delta\phi)}\, S(\Delta\eta,\Delta\phi).
  \end{equation}
  The signal pair distribution, $S(\Delta\eta,\Delta\phi)$,
  represents the \pttrk-weighted yield of jet-track pairs normalized by $N_\text{jets}$ from the same event:
  \begin{equation}
  \label{eq:signal}
  S(\Delta\eta,\Delta\phi) = \frac{1}{N_\text{jets}}\frac{\rd^{2}\Sigma \pt^\text{same}}{\rd\Delta\eta\, \rd\Delta\phi}\ .
  \end{equation}
  The mixed-event pair distribution,
  \begin{equation}
  \label{eq:background}
  ME(\Delta\eta,\Delta\phi) = \frac{1}{N_\text{jets}}\frac{\rd^{2}N^{\text{mix}}}{\rd\Delta\eta\, \rd\Delta\phi}\ ,
  \end{equation}
  is constructed to account for pair-acceptance effects, with $N^{\text{mix}}$ denoting the number of mixed-event jet-track pairs.

Signal and mixed event correlations are both corrected for tracking efficiencies on a per-track basis, using the efficiency parametrization defined as a function of centrality, \pttrk, $\eta$, $\phi$, and local charged-particle density, as in Ref.~\cite{mpt}. The ratio $ME(0,0)/ME(\Delta\eta,\Delta\phi)$ establishes the correction normalization, with $ME(0,0)$ representing the mixed-event associated yield for jet-track pairs going in approximately the same direction and thus having full pair acceptance.

\subsection{Separation of correlations into jet-peak and long-range components}

\begin{figure}[h!]
\centering

\includegraphics[width=0.9\textwidth]{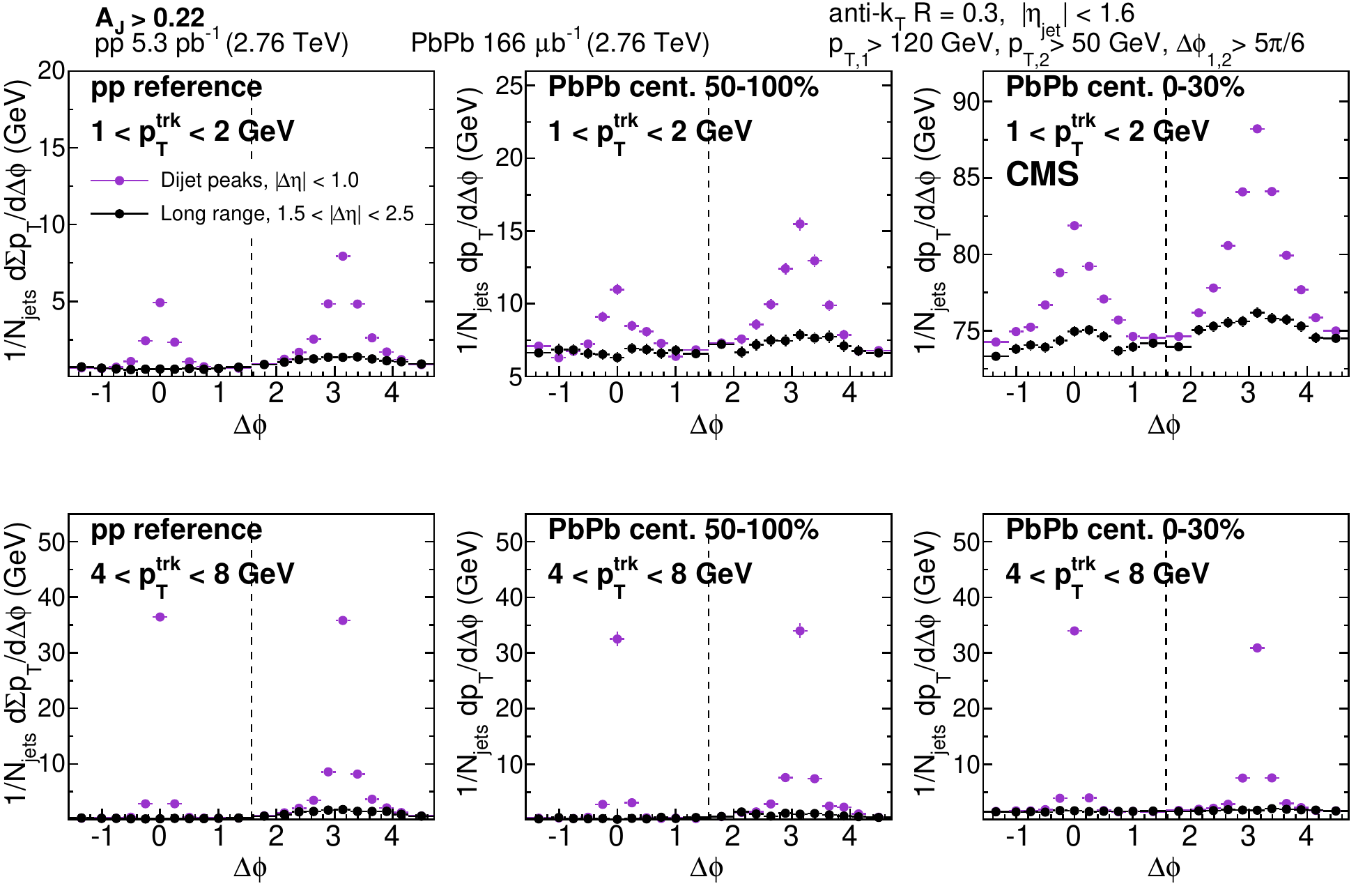}
\caption{Jet-track correlation distributions for unbalanced dijet events ($A_\mathrm{J}>0.22$), projected over $\abs{\Delta\eta}<$1. and overlaid with a long-range distribution projected over $1.5<\abs{\Delta\eta}<2.5$.  The region $\abs{\Delta\phi}<\pi/2$ is taken from the leading jet correlation, while the region $\pi/2<\Delta\phi<3\pi/2$ is taken from the subleading jet correlation.  The top row shows the low \pttrk bin $1<\pttrk<2$\GeV, while the bottom row shows the high \pttrk bin $4<\pttrk<8$\GeV.  Statistical uncertainties are shown with vertical bars.}
\label{fig:SidebandDemo}
\end{figure}

After the mixed-event correction, correlations to the leading and subleading jets show a \linebreak Gaussian-like peak confined to the region $\abs{\Delta\eta}<1.5$, on top of a significant combinatorial and long-range-correlated background.  To separate the long-range azimuthally-correlated and uncorrelated distributions under the peaks from jets, the ``sideband'' regions $1.5<|\Delta \eta|<2.5$ of both the leading and subleading jet acceptance-corrected correlations are projected into $\Delta\phi$.  Previous studies have found no $\Delta\eta$-dependence of the long-range underlying event distributions in this $\abs{\Delta\eta}$ range~\cite{Chatrchyan:2013kba,HIN_2014_016}. Figure~\ref{fig:SidebandDemo} illustrates these long-range distributions in $\Delta\phi$ for pp, most peripheral PbPb, and most central PbPb data for representative bins at low (upper panels) and high (lower panels) \pttrk of the tracks in unbalanced dijet events (with $A_\mathrm{J}>0.22$).  For illustration, the range $\abs{\Delta\phi}<\pi/2$ of the leading and subleading long-range $\Delta\phi$ distributions are shown as a combined $2\pi$ distribution, with the subleading jet distribution shifted by $\pi$ to show the full underlying event correlation with respect to the leading jet direction. The visible asymmetry in this long-range distribution in pp data is attributed to the presence of additional jets and other contributions that must, by momentum conservation, be present on the subleading side of the unbalanced dijet system.

To isolate the Gaussian-like leading and subleading jet peaks, this long-range distribution is propagated over the full range $\abs{\Delta\eta}<2.5$ and subtracted in 2D from the mixed-event-corrected signal correlation.  In addition, the leading side of the long-range distribution is subtracted from the subleading side (in the illustration shown, the distribution for $\abs{\Delta\phi}<\pi/2$ is subtracted from the distribution for $\pi/2<\Delta\phi<3\pi/2$) to obtain a measurement of subleading-to-leading asymmetry in the long-range correlated background.  With this, the three contributions to the dijet hemisphere momentum balance have been identified:  leading jet peak, subleading jet peak, and subleading-to-leading two-dimensional underlying event asymmetry.

\subsection{Corrections and systematic uncertainties}
\label{sec:systematics}

Simulation-based corrections are applied to correlations to account for two biases in jet reconstruction: a bias toward selecting jets that are found on upward fluctuations in the background (relevant for PbPb only), and a bias toward selecting jets with harder fragmentation (affecting PbPb and pp similarly). For the former, to estimate and subtract the contribution to the excess yield due to background fluctuation bias in jet reconstruction, a similar procedure to that outlined in previous CMS studies~\cite{CMS:2012vba} is followed. Simulations are performed in \PYTHIA{}+\HYDJET samples with reconstructed jets, and correlations are constructed excluding particles generated with the embedded \PYTHIA hard-scattering process.  A Gaussian fit to the excess is subtracted as a correction from the data results, and half its magnitude is assigned as the associated systematic uncertainty.

The second bias is toward the selection of jets with fewer associated tracks in both pp and PbPb data for all \pttrk selections studied, due to the fact that jets with harder fragmentation are more likely to be successfully reconstructed than jets with softer fragmentation.  Following the method used in Refs.~\cite{mpt,HIN_2014_016}, corrections are derived for this jet fragmentation function (JFF) bias and for the related possible effect of "jet swapping" between leading, subleading, and additional jets by comparing correlated per-trigger particle yields for all reconstructed jets versus all generated jets.  This correction is derived for each jet selection in a \PYTHIA-only simulation, and also in \PYTHIA{}+\HYDJET events, excluding \HYDJET tracks from the correction determination. The variation between the JFF and jet swapping correction derived from \PYTHIA embedded into \HYDJET simulation at different centralities is assigned as a systematic uncertainty in this correction.  This uncertainty is less than 2\% for all \pttrk selections, and converges to zero at high \pttrk.

Jet reconstruction-related sources of systematic uncertainty in this analysis include the two reconstruction biases as discussed above, as well as a residual JES uncertainty that accounts for possible differences of calorimeter response in data and simulation.  In simulation, for example, there is a difference in the JES between quark and gluon jets (about 2\% at 120\GeV~\cite{mpt}), meaning that medium-induced changes in jet flavor could result in either over-correction or under-correction of jet energy, and a resulting bias in jet selection.  To evaluate this residual JES uncertainty, we vary the leading jet selection threshold by 3\% to account for possible differences in data versus simulated calorimeter response. The resulting maximum variations in total correlated particle yield are found to be within 3\% in all cases, and we conservatively assign 3\% to account for this systematic uncertainty source.

The tracking efficiency correction uncertainty is estimated from the ratio of corrected reconstructed yields and generated yields in \PYTHIA and \PYTHIA{}+\HYDJET simulated events, by using generator-level charged particles as a reference.  This uncertainty is found to be $\sim2$--4\% for PbPb and pp collisions, with the greater value corresponding to a higher multiplicity and lower momentum range of selected tracks.  To account for possible track reconstruction differences in data and simulation, a residual uncertainty in track reconstruction efficiency and misidentification rate corrections is estimated to be 5\%~\cite{mpt}.

The uncertainty arising from pair acceptance effects is estimated by considering the sideband asymmetry after dividing by the mixed-event correlation.  Each sideband region of the final background-subtracted $\Delta\eta$ distribution ($-2.5<\Delta\eta<-1.5$ and $1.5<\Delta\eta<2.5$) is separately fit with a constant.  The greater of these two deviations from zero is assigned as systematic uncertainty, and is found to be within 5--9\% for the lowest \pttrk bin. The uncertainty resulting from the event decomposition is determined by evaluating point-to-point variations in the side-band projections used to estimate long-range correlation contributions. The event decomposition uncertainty is found to be within 2--5\% for 0--30\% central PbPb data in the the lowest \pttrk bin where the background is most significant compared to the signal level, and decreases for less central collisions and for higher \pttrk tracks ($\pttrk>2$\GeV).

The systematic uncertainties from the sources discussed above are added in quadrature for the final result. Table~\ref{tab:sys} lists the upper limits on the estimated contributions from the individual sources described above.

\begin{table}[htbp]
\centering
\topcaption{This table summarizes the systematic uncertainties in the measurement of the jet-track correlations in PbPb and pp collisions. Upper and lower limits
are shown as a function of collision centrality.  Upper values correspond to the uncertainties at lowest \pttrk.}
\label{tab:sys}
\begin{tabular}{lccccc}
\hline
\multicolumn{1}{c}{Source} & 0--30\% &   30--50\% & 50--100\% & pp \\
\hline\\[-0.2ex]
Balanced jet selection ($A_\mathrm{J} < 0.22$):\\
\hline
Background fluctuations                       & 1--8\%  & 1--3\% & 0--1\% & --- \\
JFF bias and jet swapping               & 0--2\%  &   0--2\% & 0--2\% &    0--2\% \\
Residual JES                                   & 3\% &      3\% &    3\% &    3\% \\
Tracking efficiency         & 4\%  & 4\% & 4\% & 3 \%  \\
Residual track efficiency corr.              &    5\%  &    5\% &    5\% &    5\% \\
Pair acceptance corrections                  & 5--9\%  & 4--8\% & 2--6\% & 2--3\% \\
Event decomposition                      & 2--5\% & 2--5\% & 2--5\% & 1--2\% \\
\hline
\multicolumn{1}{c}{Total}                                        & 9--15\%  & 8--13\% & 8--10\% & 7--8\% \\
\hline\\[-0.2ex]
Unbalanced jet selection ($A_\mathrm{J} > 0.22$):\\
\hline
Background fluctuations                       & 1--10\% &  1--5\% & 0--2\% & --- \\
JFF bias and jet swapping               & 0--2\% &      0--2\% & 0--2\% &    0--2\% \\
Residual JES                                   & 3\% &     3\% &    3\% &    3\% \\
Tracking efficiency           & 4\% & 4\% & 4\% & 3 \%  \\
Residual track efficiency corr.              &    5\% &    5\% &    5\% &    5\% \\
Pair acceptance corrections                  & 5--9\%  & 4--8\% & 2--6\% & 2--3\% \\
Event decomposition                    & 2--5\% &  2--5\% & 2--5\% & 1--2\% \\
\hline
\multicolumn{1}{c}{Total}                                        & 9--16\%  & 8--13\% & 8--10\% & 7--8\% \\
\hline

\end{tabular}
\end{table}
\section{Results}

In this analysis, two-dimensional $\Delta\eta$--$\Delta\phi$ momentum distributions with respect to high-\pt leading and subleading jets are studied differentially in centrality and \pttrk.   First, the jet momentum density profile is measured as a function of $\Delta r = \sqrt{(\Delta\eta)^{2}+(\Delta\phi)^{2}}$ for all dijet events, comparing PbPb to pp jet shapes up to $\Delta r = 1$.  Next, the sample is divided into balanced ($A_\mathrm{J}<0.22$) and unbalanced ($A_\mathrm{J}>0.22$) dijet events, and the overall subleading-to-leading hemisphere momentum balance is evaluated by subtracting the distribution of \pttrk about the leading jet from the distribution of \pttrk about the subleading jet .  This overall hemisphere momentum balance is then further decomposed into contributions from the leading and subleading jet peaks and from the underlying event-wide long-range asymmetry.  The jet peak shapes are quite similar in $\Delta\eta$ and $\Delta\phi$, as reported in Ref.~\cite{HIN_2014_016}, but the long-range ``ridge-like'' distribution is independent of $\Delta\eta$ within the fiducial acceptance and uncertainties, while showing a clear $\Delta\phi$ dependence, as is visible in Fig.~\ref{fig:SidebandDemo}. To avoid convolving these two different trends, we present the results that follow as a function of $\Delta\phi$, including  the overall subleading-to-leading hemisphere momentum balance and respective contributions from the leading and subleading peaks and underlying long-range asymmetry as a function of $\Delta\phi$. Finally, we  summarize our findings  by presenting the total momentum in each \pttrk bin, integrated over each hemisphere.

\subsection{Measurement of radial jet momentum density profile}

After subtraction of the long-range background, \pttrk correlations in $\Delta\eta$ and $\Delta\phi$ may be used to obtain measurements of jet shape as a function of $\Delta r$ by direct integration.  Jet shape $\rho(\Delta r)$ is defined as:
\begin{equation}
\rho(\Delta r) = \frac{1}{\delta r} \frac{1}{N_\text{jets}} \large{\Sigma_\text{jets}} \frac{\Sigma_{\text{tracks} \in (r_{a},r_{b})} \pttrk}{\pt^\text{jets}}.
\end{equation}

\begin{figure}[h!]
\centering
\includegraphics[width=0.99\textwidth]{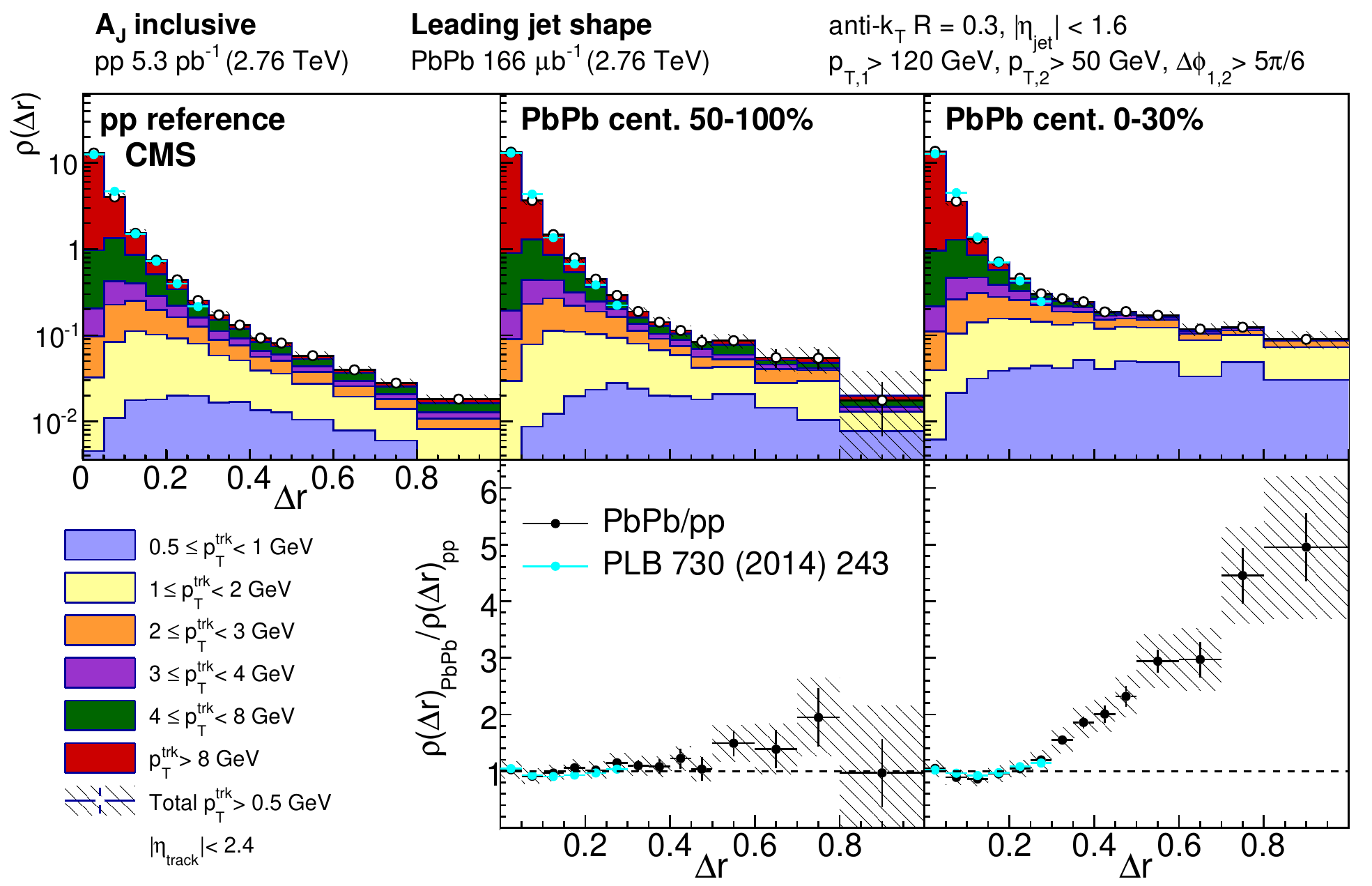}
\caption{Top row:  leading jet shape $\rho(\Delta r)$ for the pp reference, and peripheral and central PbPb data, shown for all tracks with $\pt > 0.5$\GeV and decomposed by \pttrk (with \pttrk ranges denoted by different color shading).  Shapes are normalized to unity over the region $\Delta r<0.3$ for comparison with the published reference shown (Ref.~\cite{Chatrchyan:2013kwa}).  Bottom row:  leading jet shape ratio $\rho(\Delta r)_\mathrm{PbPb}/\rho(\Delta r)_{\Pp\Pp}$, again with the published reference.  Statistical uncertainties are shown with vertical bars, and systematic uncertainties are shown with shaded boxes.}
\label{fig:JetShape_Leading}
\end{figure}

The jet shape $\rho(\Delta r)$ is extracted by integrating 2D jet-peak momentum distributions in annuli with radial width $\delta r = 0.05$, where each has an inner radius of $r_{a} = r - \delta r/2$ and outer radius of $r_{b} = r + \delta r/2$.  A previous CMS study measured the jet shape $\rho(\Delta r)$ within the jet cone radius $\Delta r = 0.3$~\cite{Chatrchyan:2013kwa}; for comparison with this previous result, distributions are normalized to integrate to unity within the radius $\Delta r < 0.3$.  In Fig.~\ref{fig:JetShape_Leading}, the leading jet shape measured with this correlation technique is compared to the published CMS measurement~\cite{Chatrchyan:2013kwa}.  The new jet shape measurement is performed differentially in \pttrk, in bins ranging from $0.5 < \pttrk < 1$\GeV to $\pttrk > 8$\GeV.  With the advantages provided by the correlation technique, the radial jet momentum density profile measurements are also extended in this analysis to $\Delta r=1$. The $\Delta r=1$ limit is driven by the pp data, which has no correlated yields within our sample at larger $\Delta r$.  The leading jet shape is found to be very similar to that in the previous measurement for an inclusive jet selection of all jets with $\pt>100$\GeV, despite small differences in the jet selection.

\begin{figure}[h!]
\centering
\includegraphics[width=0.99\textwidth]{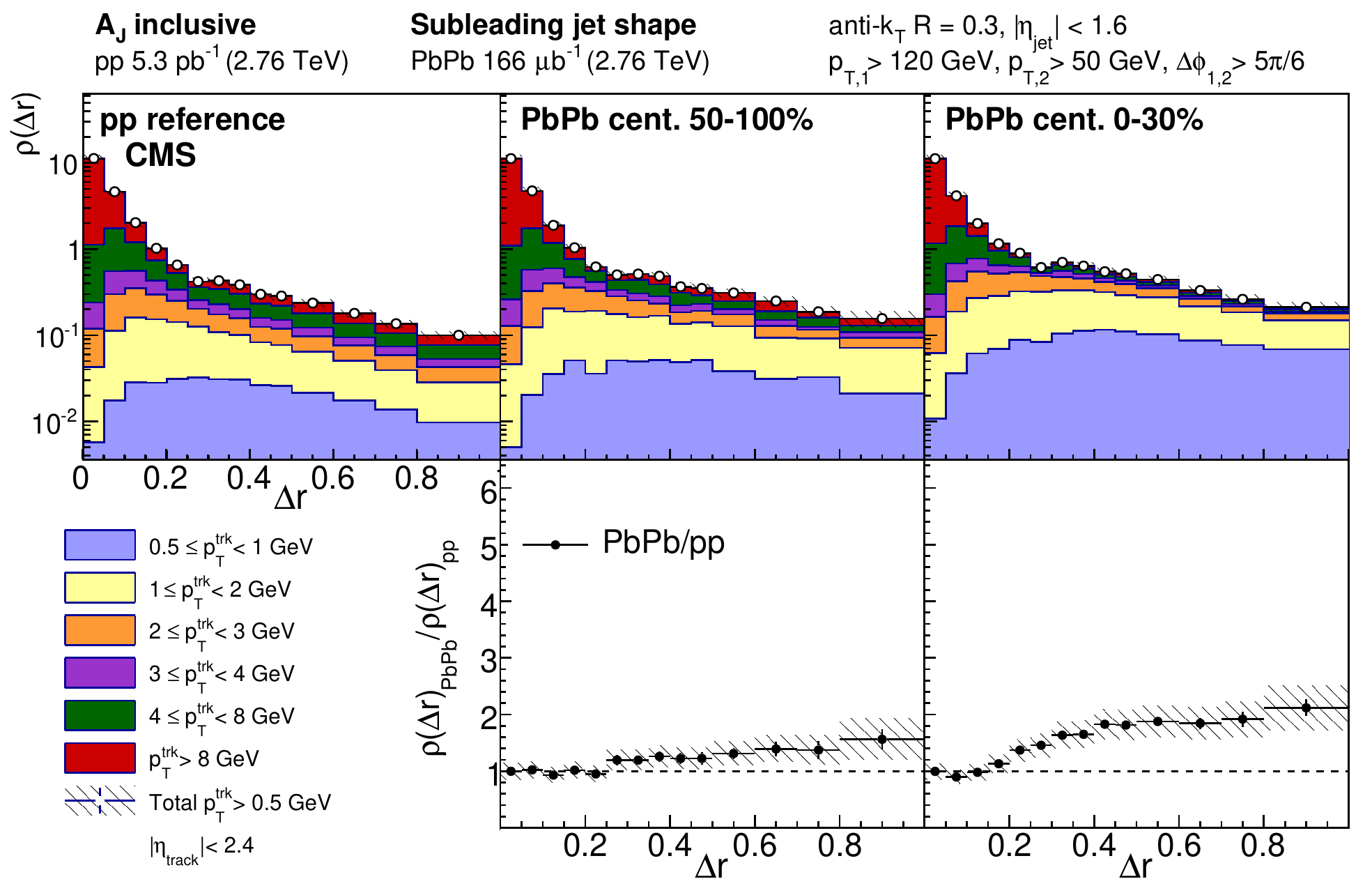}
\caption{Top row:  subleading jet shape $\rho(\Delta r)$ for pp reference and peripheral and central PbPb data, shown for all tracks with $\pt> 0.5$\GeV and decomposed by \pttrk (with \pttrk ranges denoted by different color shading), normalized to unity over the region $\Delta r<0.3$.  Bottom row:  subleading jet shape ratio $\rho(\Delta r)_\mathrm{PbPb}/\rho(\Delta r)_{\Pp\Pp}$.   Statistical uncertainties are shown with vertical bars, and systematic uncertainties are shown with shaded boxes.}
\label{fig:JetShape_SubLeading}
\end{figure}

A new measurement of subleading jet shape is presented in Fig.~\ref{fig:JetShape_SubLeading}.  Significant jet shape modifications are evident in central PbPb events for both leading and subleading jets with respect to the pp reference measurement, while in peripheral PbPb events the jet shapes are similar to the pp reference. Broadening of the jet structure is an expected consequence of jet quenching in theoretical models~\cite{Qin:2015srf}. Here it is important to note that the broadening in central PbPb collisions is relative to the jets of the same type (leading or subleading) in pp collisions.  The subleading jets are of lower \pt by selection in both PbPb and pp, and significantly broader than leading jets in pp data. Thus, although subleading jets in central PbPb collisions are softer and broader than leading jets in these collisions, the relative jet shape modification (expressed via a ratio to pp data) is greater for leading jets since these are compared to the narrower pp leading jets.

\subsection{Azimuthal distribution of charged-particle transverse momentum}

\begin{figure}[h!]
\centering
\includegraphics[width=0.99\textwidth]{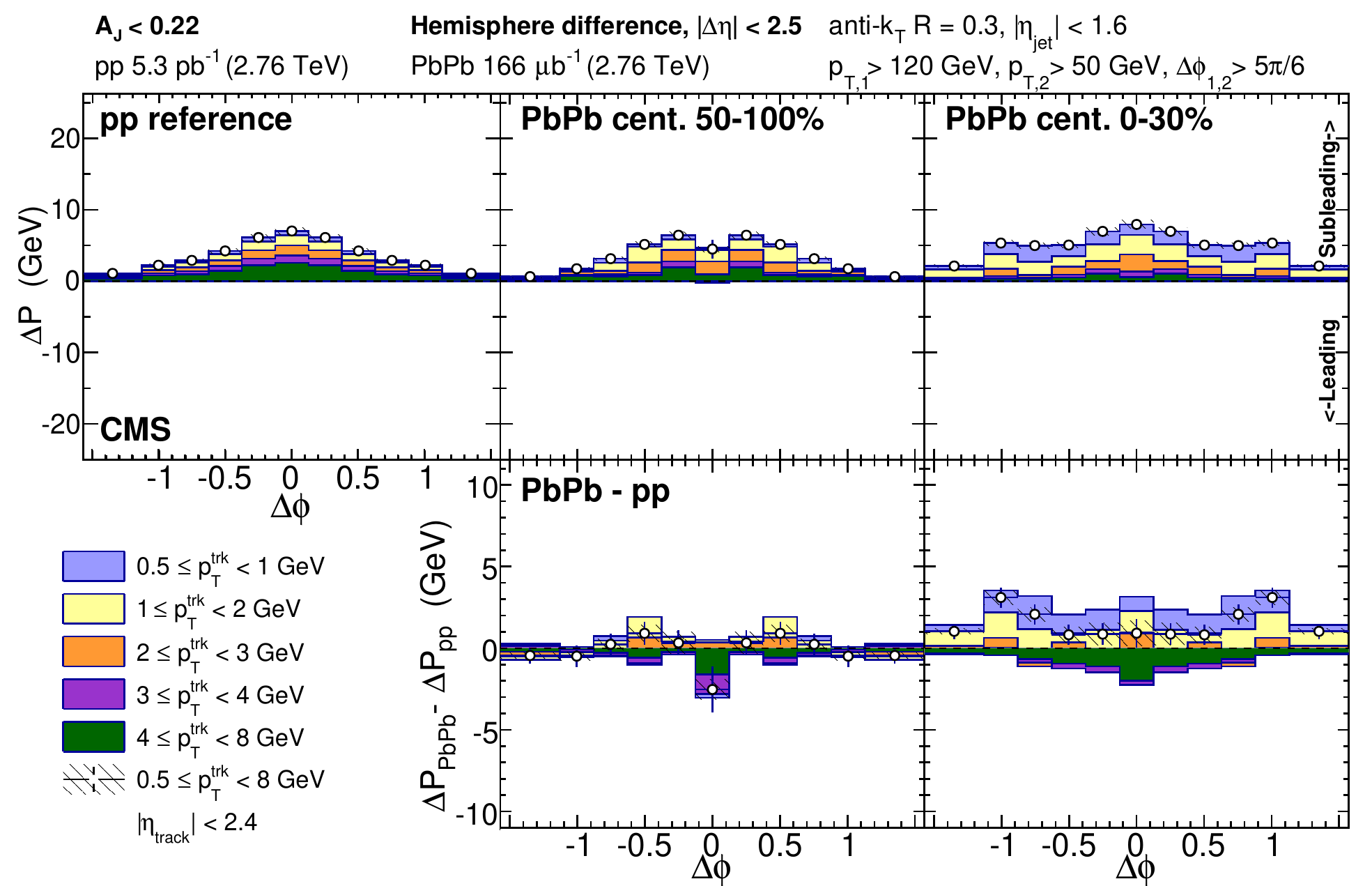}
\caption{Top row:  difference of the total \pttrk distributions between  subleading and leading jet hemispheres, projected on $\Delta\phi$,  for balanced dijet events with $A_\mathrm{J} < 0.22$ shown differentially by \pttrk for pp reference, peripheral PbPb, and central PbPb data.  Bottom row:  PbPb--pp difference in these $\Delta\phi$ momentum distributions.  Statistical uncertainties are shown with vertical bars, and systematic uncertainties are shown with shaded boxes.}
\label{fig:MpT_nobkgsub_Aj0_Aj22}
\end{figure}

\begin{figure}[h!]
\centering
\includegraphics[width=0.99\textwidth]{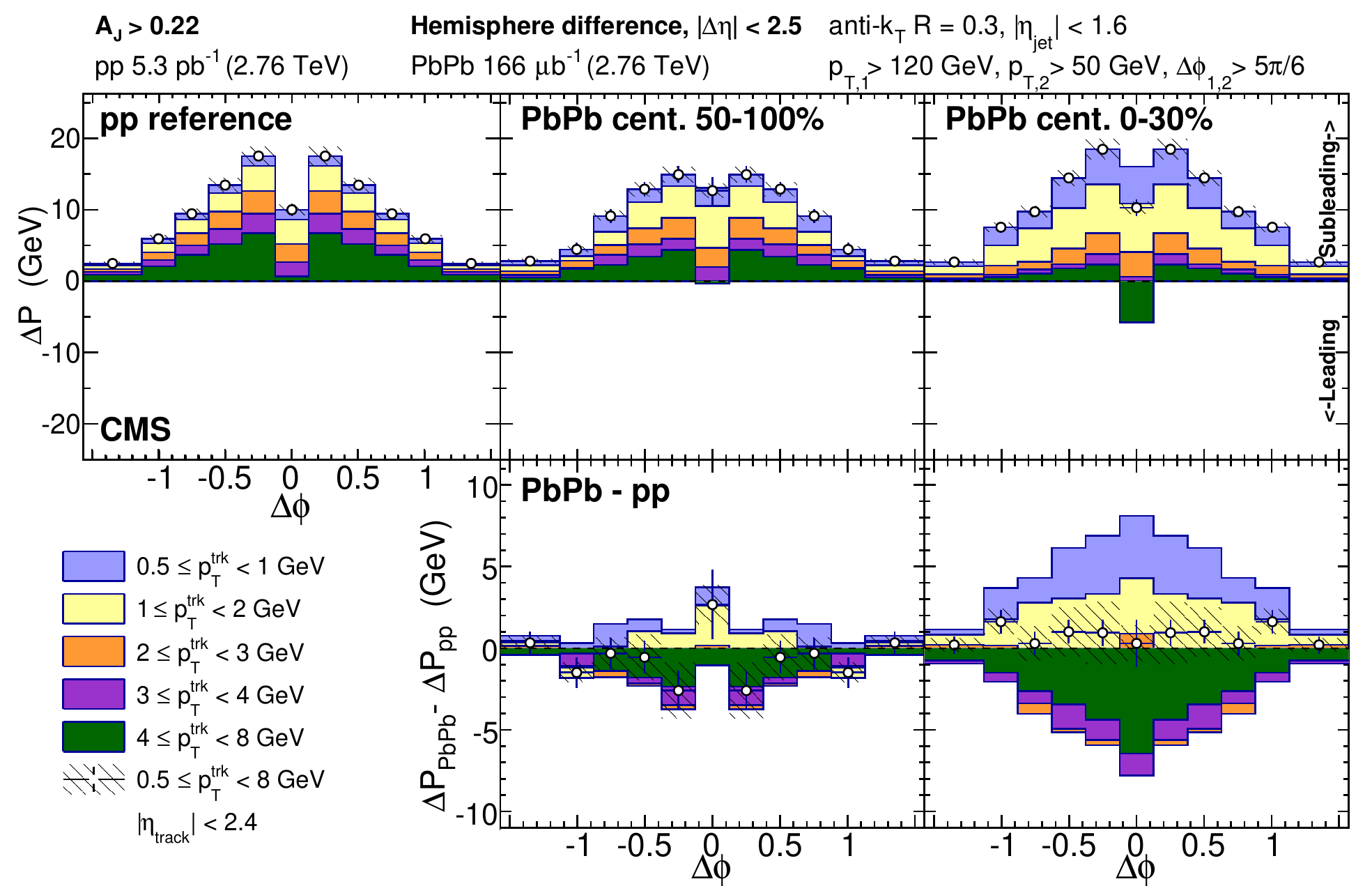}
\caption{Top row:  difference of the total \pttrk distributions between  subleading and leading jet hemispheres, projected on $\Delta\phi$,  for unbalanced dijet events with $A_\mathrm{J} > 0.22$, shown differentially by \pttrk for the pp reference data, peripheral PbPb, and central PbPb data.  Bottom row:  PbPb--pp difference in these $\Delta\phi$ momentum distributions.  Statistical uncertainties are shown with vertical bars, and systematic uncertainties are shown with shaded boxes.}
\label{fig:MpT_nobkgsub_Aj22_Aj75}
\end{figure}

To investigate in detail how the modification of the jet peaks contributes to the overall redistribution of \pttrk flow reported in~\cite{mpt}, we measure the transverse momentum balance in the event via $\Delta\eta$--$\Delta\phi$ \pttrk correlations to subleading and leading jets.  First, we present the hemisphere-wide balancing distribution of \pttrk around the subleading versus the leading jets in Figs.~\ref{fig:MpT_nobkgsub_Aj0_Aj22} and~\ref{fig:MpT_nobkgsub_Aj22_Aj75}, for balanced and unbalanced dijet events, respectively.  These figures show the per-event azimuthal distribution of \pttrk about the leading or subleading jet axis, denoted
$\mathrm{P} = 1/N_\text{evt}\,\rd\Sigma \pt/\rd\Delta\phi$, with the subleading-to-leading hemisphere difference of this distribution denoted $\Delta \mathrm{P}$.  For display, all distributions are symmetrized in $\Delta\phi$.  For both balanced and unbalanced dijet events, a broad excess of soft particles is evident in the subleading versus leading hemisphere in central PbPb collisions relative to the pp reference data.  This reflects the greater quenching of the subleading jet.  In the unbalanced selection, as required by the momentum conservation, the signal is enhanced in both pp and PbPb data:  in pp data a large excess of particles with $\pttrk>3$\GeV is present on the subleading side.  This excess compensates for the smaller contribution of the highest \pt particles in the jet itself.  In peripheral PbPb data the distribution is quite similar to the pp reference, while in central PbPb data this balancing distribution consists mostly of soft particles with $\pttrk<3$\GeV, consistent with the findings of the previous CMS study~\cite{mpt}.  To demonstrate these medium modifications more clearly, the difference in yield between PbPb and pp collisions is shown in the bottom panels of Figs.~\ref{fig:MpT_nobkgsub_Aj0_Aj22} and~\ref{fig:MpT_nobkgsub_Aj22_Aj75}.  For presentation, tracks with $\pttrk>8$\GeV are not included in these figures, so that it is possible to zoom in on the low-\pttrk structures and modifications.

\begin{figure}[h!]
\centering
\includegraphics[width=0.99\textwidth]{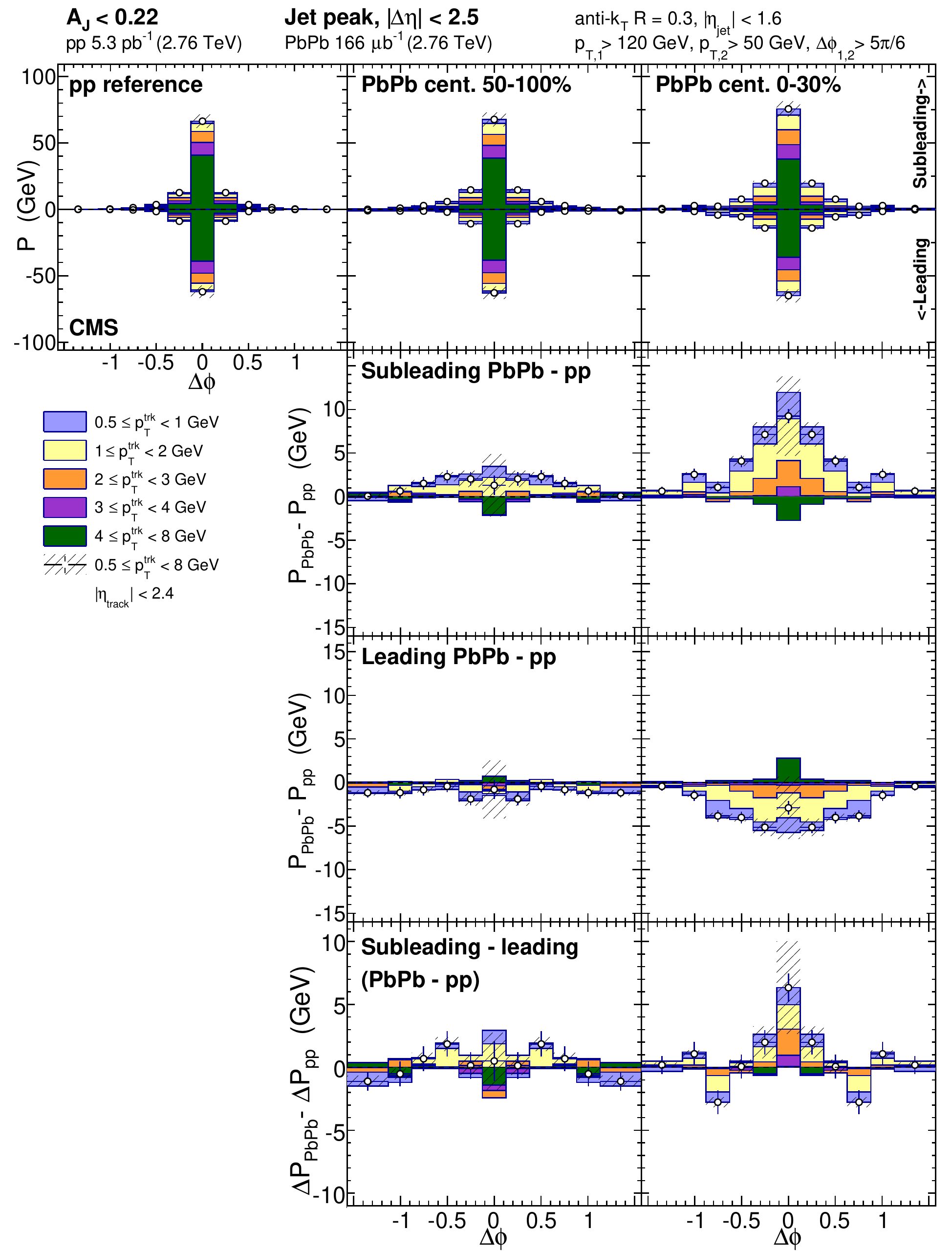}
\caption{Top row:  jet-peak (long-range subtracted) distribution in $\Delta\phi$ of \pttrk about the subleading (plotted positive) and leading (plotted negative) jets for balanced dijet events with $A_\mathrm{J} < 0.22$.  Middle rows:  PbPb--pp momentum distribution differences for subleading and leading jets.  Bottom row:  PbPb--pp, subleading--leading double difference in these $\Delta\phi$ momentum distributions.  Statistical uncertainties are shown with vertical bars, and systematic uncertainties are shown with shaded boxes.}
\label{fig:MpT_jetrelated_Aj0_Aj22}
\end{figure}

\begin{figure}[h!]
\centering
\includegraphics[width=0.99\textwidth]{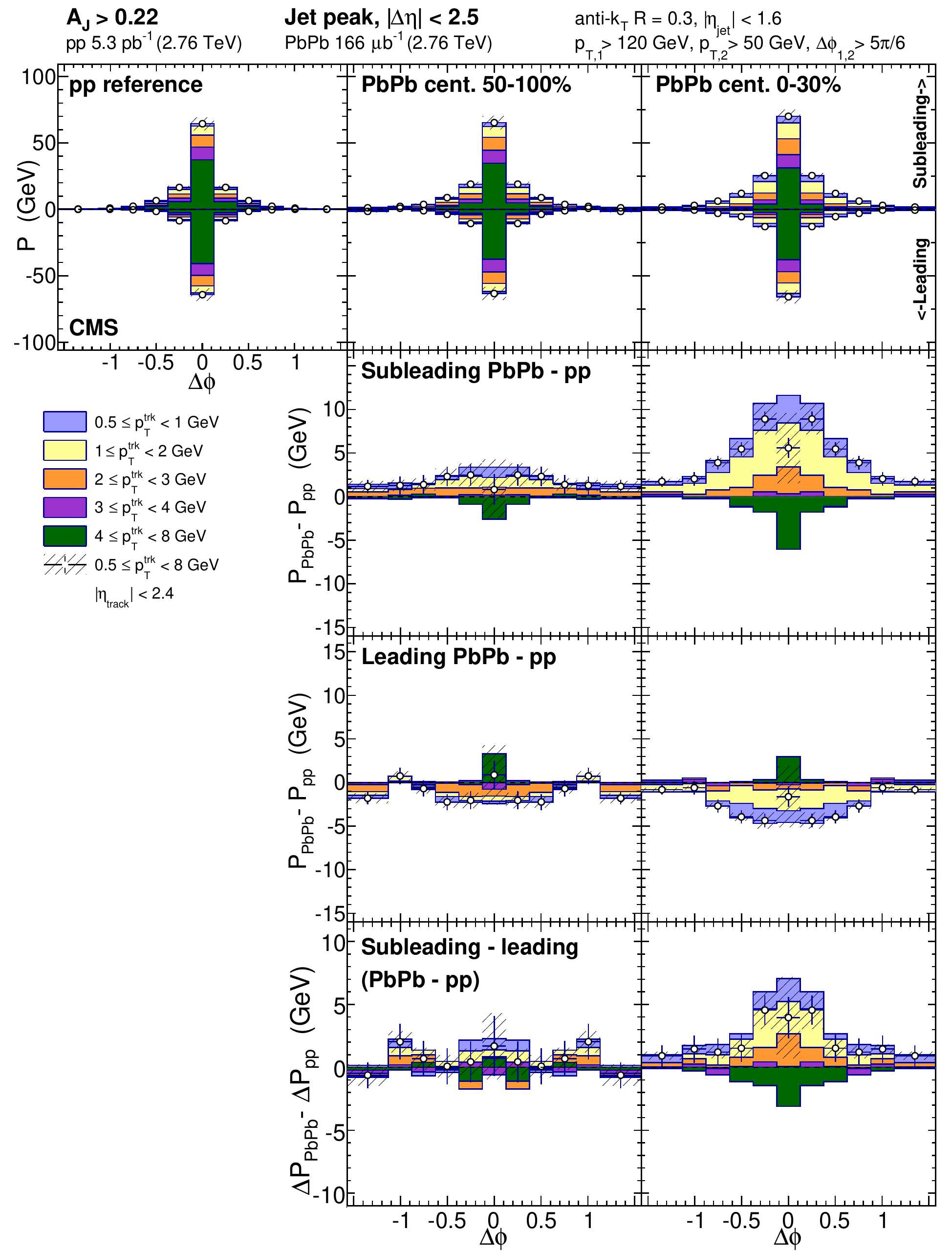}
\caption{Top row:  jet-peak (long-range subtracted) distribution in $\Delta\phi$ of \pttrk about the subleading (plotted positive) and leading (plotted negative) jets for unbalanced dijet events with $A_\mathrm{J} > 0.22$.  Middle rows:  PbPb--pp momentum distribution differences for subleading and leading jets.  Bottom row:  PbPb--pp, subleading--leading double difference in these $\Delta\phi$ momentum distributions.  Statistical uncertainties are shown with vertical bars, and systematic uncertainties are shown with shaded boxes.}
\label{fig:MpT_jetrelated_Aj22_Aj75}
\end{figure}

To elucidate the redistribution of \pttrk within the QGP, the distributions are separated into three components as stated above: the two Gaussian-like peaks about the leading and subleading jet axes, and a third component accounting for overall subleading-to-leading hemisphere asymmetry of the long-range side-band distributions (measured in the region $1.5<\abs{\Delta\eta}<2.5$).  In Figs.~\ref{fig:MpT_jetrelated_Aj0_Aj22} and~\ref{fig:MpT_jetrelated_Aj22_Aj75}, the jet peak components are shown for balanced and unbalanced jets, respectively, presenting subleading results as positive and leading results as negative (in line with the hemisphere difference measurements in Figs.~\ref{fig:MpT_nobkgsub_Aj0_Aj22} and~\ref{fig:MpT_nobkgsub_Aj22_Aj75}).  Jet peak distributions after decomposition are projected over the full range $\abs{\Delta\eta}<2.5$, again for consistency with the hemisphere difference measurements.  The top row of each panel first shows the overall distribution of momentum carried by particles with $\pttrk < 8$\GeV associated with the jet peak.  The middle two panels then assess modifications to the subleading and leading jets for each \pttrk bin.  Here, again, there is evidence of quenching of both the subleading and leading jets in central PbPb collisions relative to the pp reference data.  There is an excess of low-\pt particles correlated with the leading and subleading jet axes in both the balanced and unbalanced dijet selections, in agreement with results presented in the CMS study~\cite{HIN_2014_016}.  In unbalanced dijet events, this enhancement of soft particles turns into a depletion at higher \pttrk, and is greater on the subleading than the leading side.  The pp subleading jet peak is broader than the pp leading jet peak and, while both subleading and leading jet peaks are broader in central PbPb than pp data, the PbPb--pp excess is wider on the leading than on the subleading side.  To assess the jet peak contributions to the overall hemisphere momentum balance, the double-differential (PbPb--pp, subleading--leading) result is presented in the bottom panel.  Here it is evident that the low-\pttrk excess in central PbPb collisions is greater on the subleading than the leading side of the dijet system, but the larger subleading-to-leading excess only accounts for a portion of the total momentum redistribution in unbalanced dijet events. It is also clear that the high-\pttrk large-angle depletion observed in the overall hemisphere momentum balance distribution is not produced by the Gaussian-like jet peaks.

\begin{figure}[h!]
\centering
\includegraphics[width=0.99\textwidth]{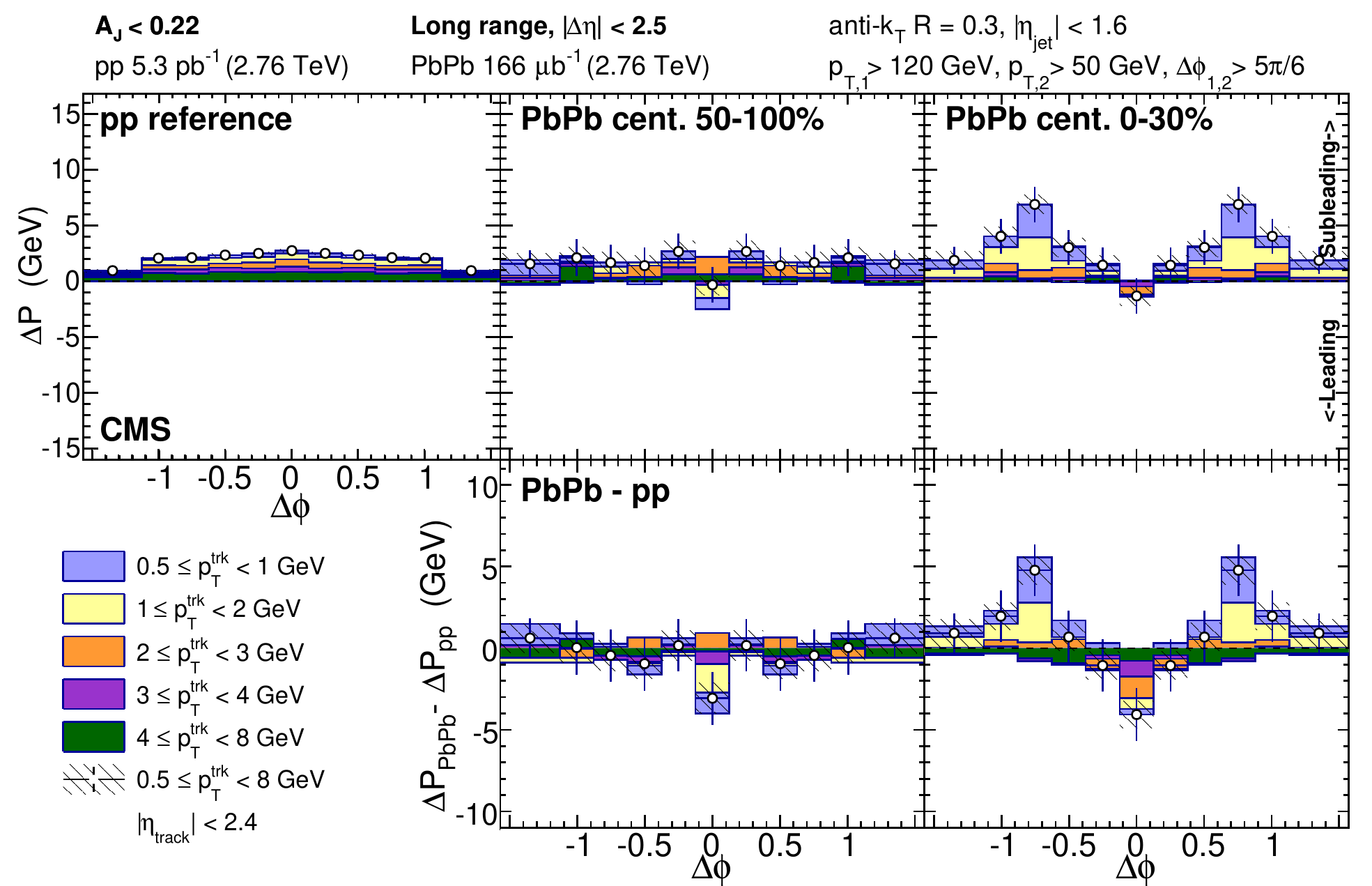}
\caption{Top row:  long-range distribution in $\Delta\phi$ of excess \pttrk in the subleading relative to leading sides for balanced dijet events with $A_\mathrm{J} < 0.22$.  Bottom row:  PbPb--pp difference in these $\Delta\phi$ long-range momentum distributions.  Statistical uncertainties are shown with vertical bars, and systematic uncertainties are shown with shaded boxes.}
\label{fig:MpT_longrange_Aj0_Aj22}
\end{figure}

\begin{figure}[h!]
\centering
\includegraphics[width=0.99\textwidth]{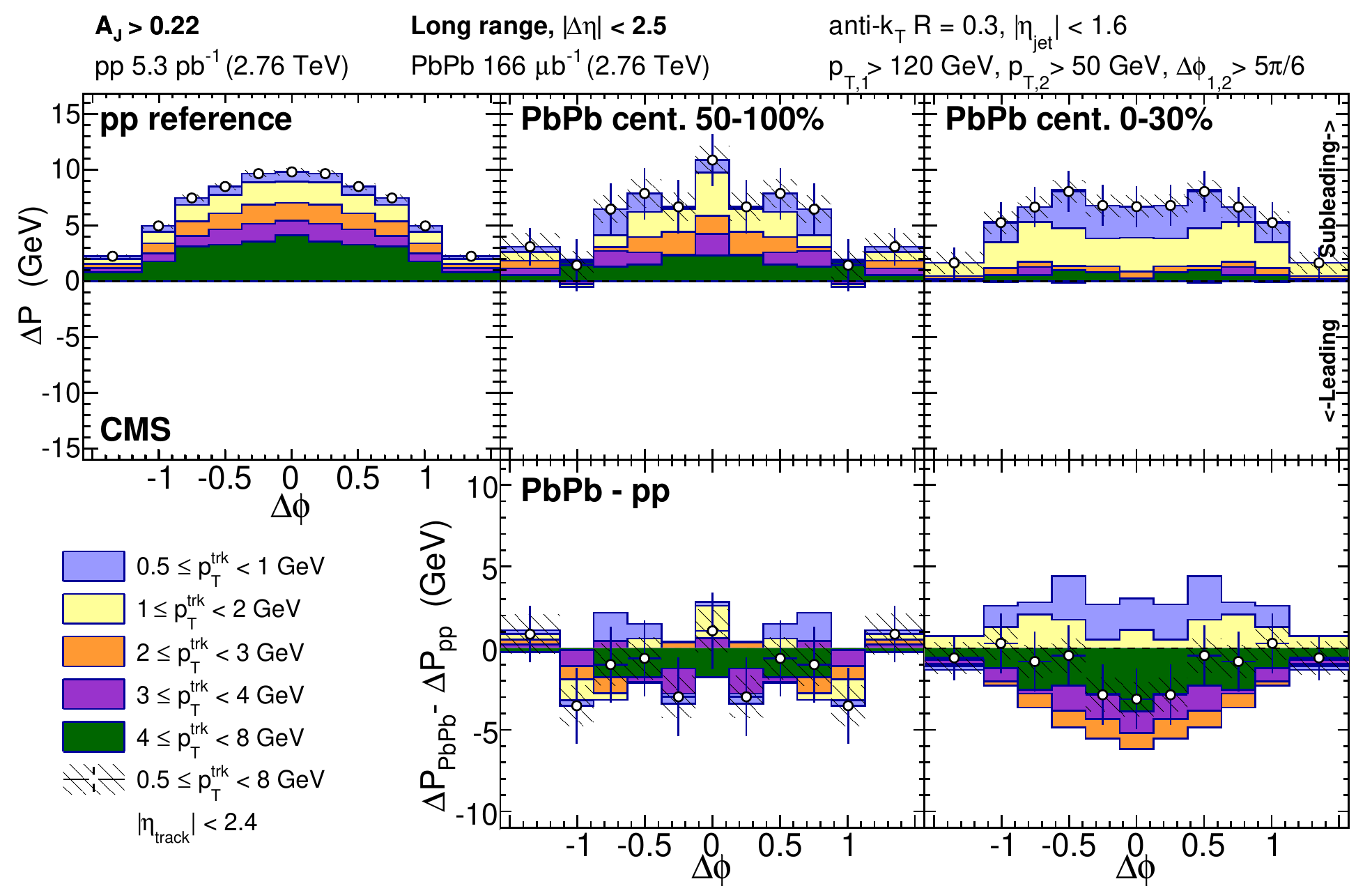}
\caption{Top row:  long-range distribution in $\Delta\phi$ of excess \pttrk in the subleading relative to leading sides for unbalanced dijet events with $A_\mathrm{J} > 0.22$.  Bottom row:  PbPb--pp difference in these $\Delta\phi$ long-range momentum distributions.  Statistical uncertainties are shown with vertical bars, and systematic uncertainties are shown with shaded boxes.}
\label{fig:MpT_longrange_Aj22_Aj75}
\end{figure}

To uncover the missing part of the total transverse momentum balance, these jet-related studies are complemented by an analysis of the long-range subleading-to-leading hemisphere asymmetry, presented in Figs.~\ref{fig:MpT_longrange_Aj0_Aj22}  and~\ref{fig:MpT_longrange_Aj22_Aj75} for balanced and unbalanced jets, respectively.  The long-range correlated background in balanced dijet events is approximately symmetric in pp and peripheral PbPb data, while in central PbPb data there is a small excess of low-\pt particles.  In unbalanced dijet events, however, there is already significant asymmetry in the pp reference data, with a large correlated excess of particles in all \pttrk classes less than 8\GeV on the subleading relative to the leading side of the underlying event.  This asymmetry reflects the presence of other hard-scattering products in the subleading hemisphere of dijet events (e.g. additional jets), as required by the momentum conservation for asymmetric dijet events in pp collisions.  In the presence of the strongly interacting medium, however, this underlying event asymmetry in asymmetric dijet events changes notably.  In peripheral PbPb collisions, an onset of some depletion of momentum carried by high-\pt particles can be seen, and in central PbPb data, subleading-to-leading underlying event excesses with $\pttrk > 2$\GeV nearly vanish.  The significance of the contribution of this long-range asymmetry to the total hemisphere imbalance is further assessed by the double difference (PbPb--pp, subleading--leading), which  is shown on the bottom panel.  The absence of a high-\pttrk component in the long-range part of the correlation suggests that events containing additional jets constitute a smaller fraction of unbalanced dijet events in PbPb than pp data.  A likely explanation for this effect is that while momentum conservation requires the presence of additional jets in unbalanced pp dijet events, in PbPb data the sample of unbalanced dijet events also includes events in which the dijet asymmetry is due to the greater quenching of the subleading jet.

\clearpage

\subsection{Integrated hemisphere momentum balance}

\begin{figure}[h!]
\centering
\includegraphics[width=0.99\textwidth]{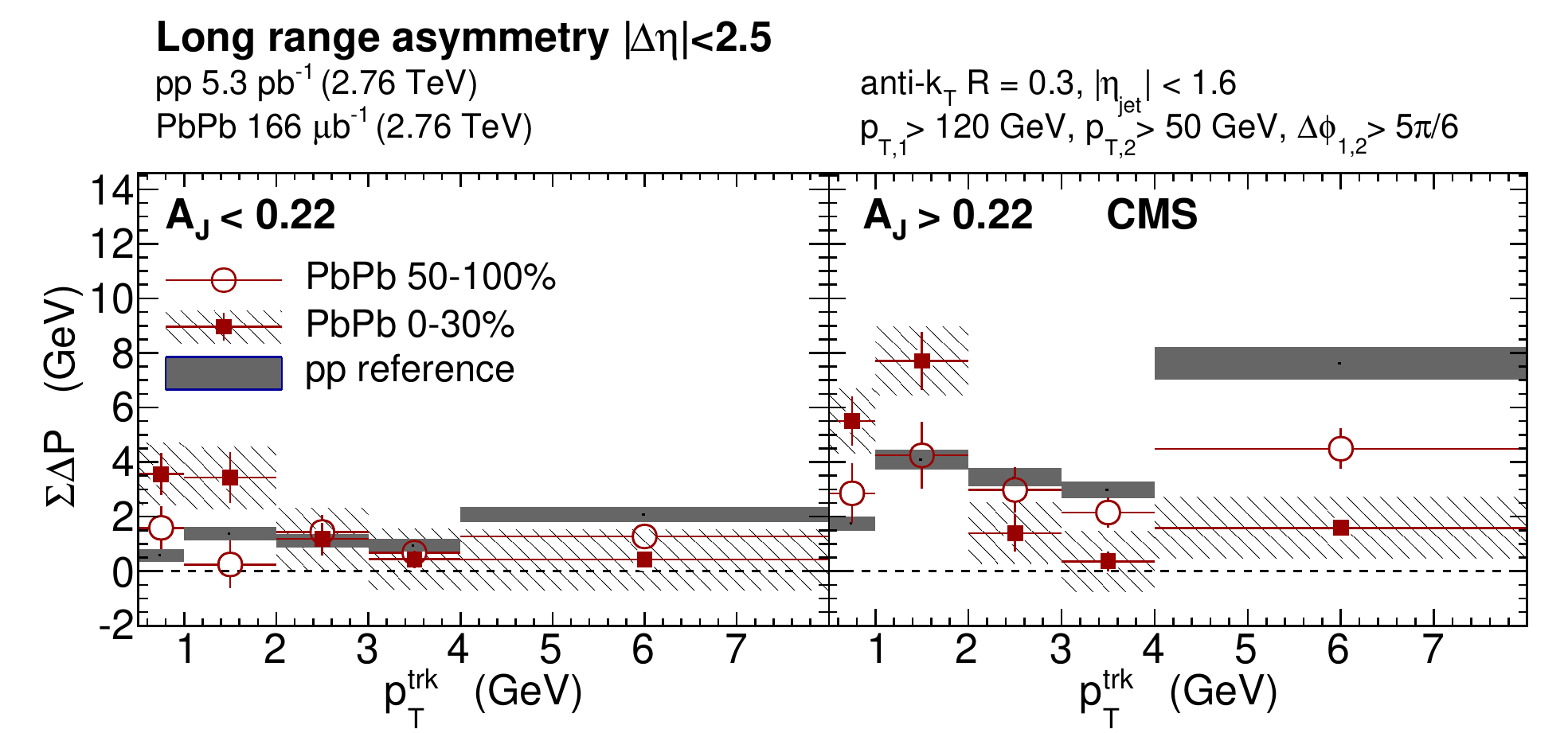}
\caption{Integrated \pt in the long-range $\Delta\phi$-correlated distribution as a function of track-\pttrk integrated over $\abs{\Delta\phi} < \pi/2$  and $\abs{\Delta\eta} < 2.5$ for pp reference, peripheral PbPb and central PbPb data for balanced compared to unbalanced dijet events.  Statistical uncertainties are shown with vertical bars, and systematic uncertainties are shown with shaded boxes.}
\label{fig:Integral_Longrange}
\end{figure}

\begin{figure}[h!]
\centering
\includegraphics[width=0.99\textwidth]{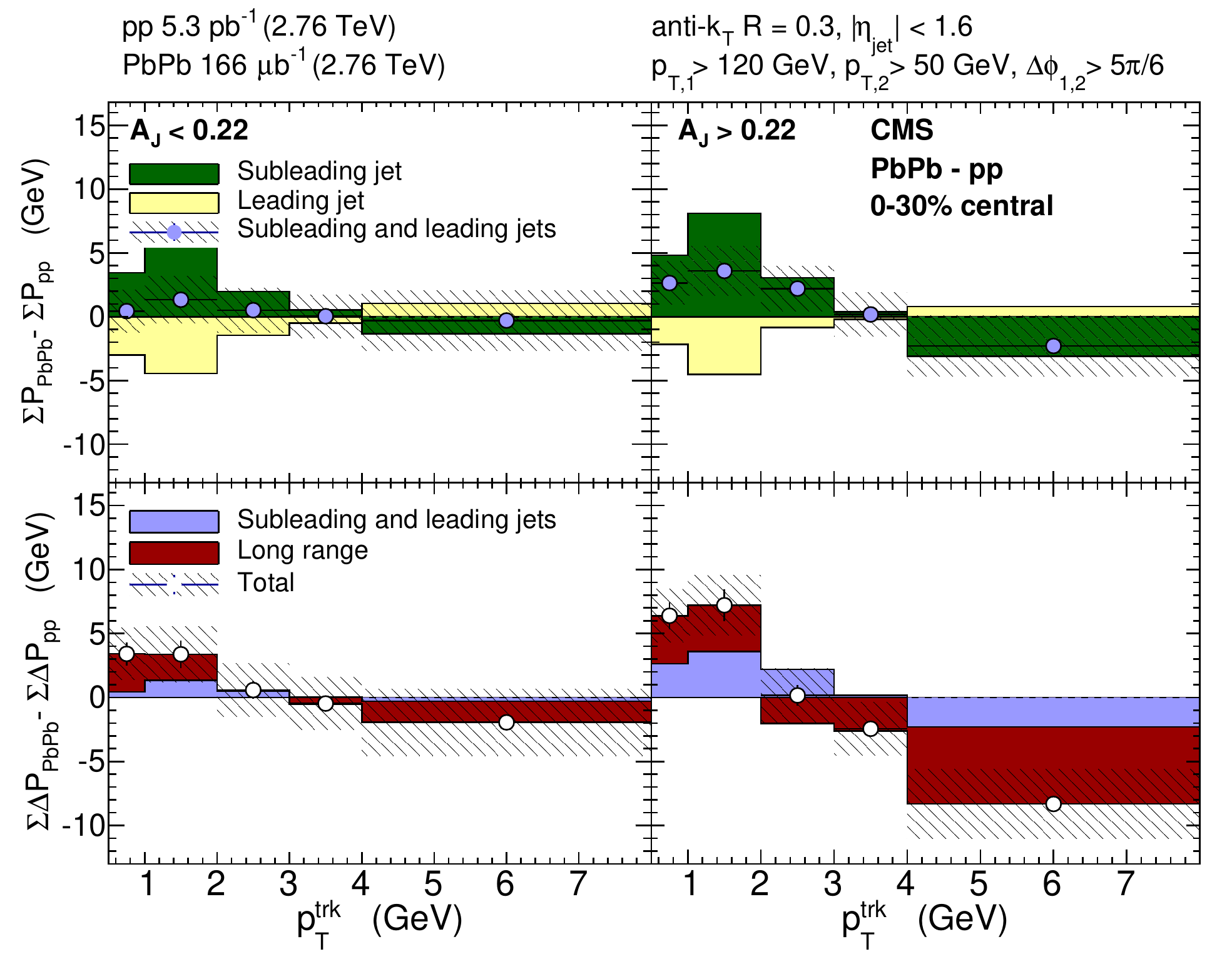}
\caption{Modifications of jet-track correlated \pttrk in central PbPb collisions with respect to the pp reference data, integrated over $\abs{\Delta\phi} < \pi/2$ and $\abs{\Delta\eta} < 2.5$.  Top row:  subleading and leading jet peaks PbPb--pp.  Bottom row:  relative contributions from jet peaks and long-range asymmetry to the double difference (PbPb--pp, subleading--leading) in total hemisphere \pttrk.  Statistical uncertainties are shown with vertical bars, and systematic uncertainties are shown with shaded boxes.}
\label{fig:Integral_Summary}
\end{figure}

\begin{figure}[h!]
\centering
\includegraphics[width=0.99\textwidth]{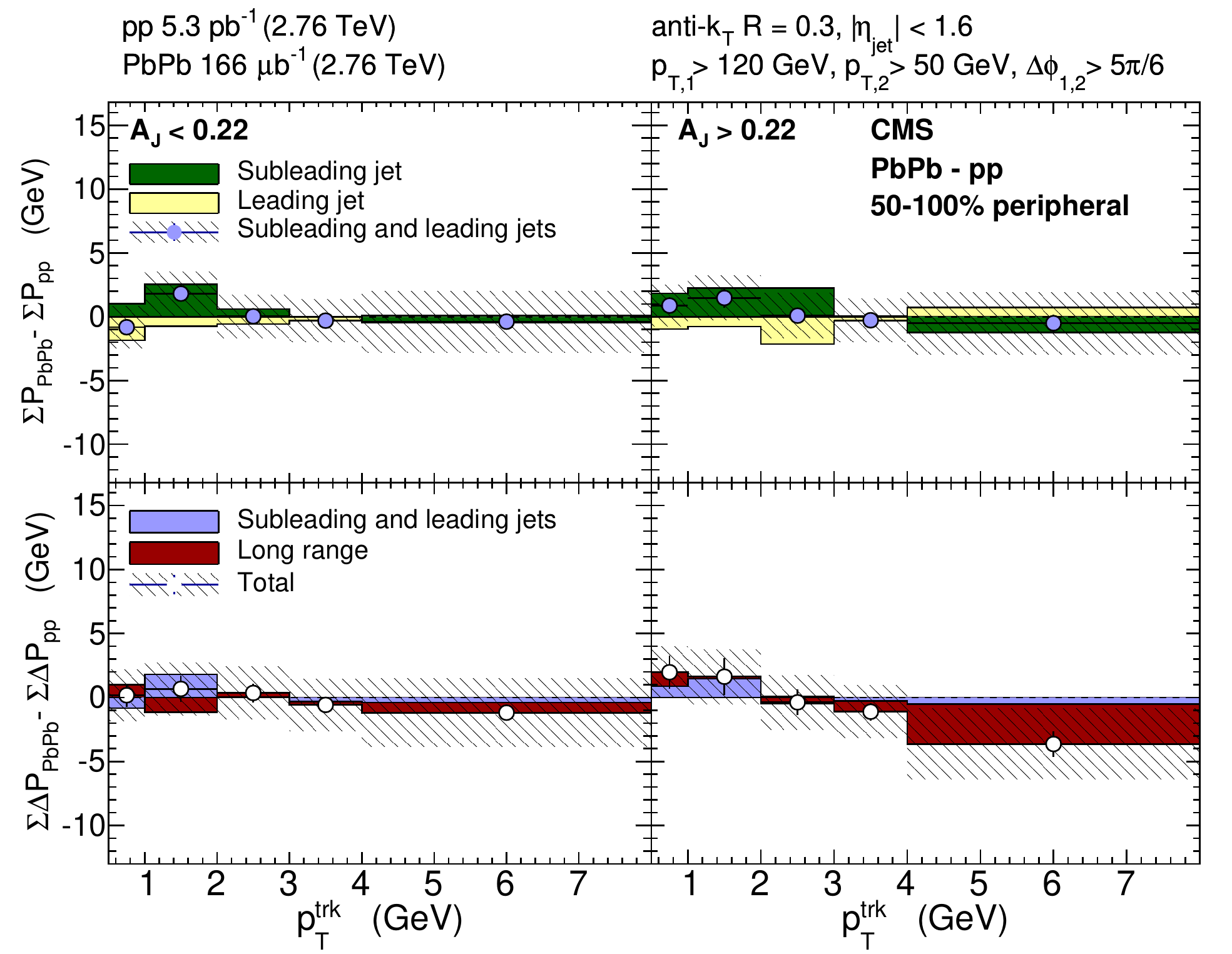}
\caption{Modifications of jet-track correlated \pttrk in peripheral PbPb collisions with respect to the pp reference data, integrated over $\abs{\Delta\phi} < \pi/2$ and $\abs{\Delta\eta} < 2.5$.  Top row:  subleading and leading jet peaks PbPb--pp.  Bottom row:  relative contributions from jet peaks and two-dimensional asymmetry to the double difference PbPb--pp, subleading--leading in total hemisphere \pttrk.  Statistical uncertainties are shown with vertical bars, and systematic uncertainties are shown with shaded boxes.}
\label{fig:Integral_Summary_Peripheral}
\end{figure}

To summarize all contributions to the overall \pttrk flow in the dijet events, we first present the hemisphere integral $\Sigma\Delta$P (where $\Delta$P represents the subleading-to-leading difference in the $\Delta\phi$ distribution of \pttrk in the event) for  this long-range asymmetry, measured in $\abs{\Delta\phi}<\pi/2$ and $\abs{\Delta\eta}<2.5$,  in Fig.~\ref{fig:Integral_Longrange}.  For balanced dijet events, the PbPb and pp integrals follow similar \pttrk dependences.
For unbalanced dijet events, the overall asymmetry rises with \pttrk in the pp reference data, but falls with \pttrk in central PbPb data, indicating the presence of different sources for the long-range momentum correlations.  Finally,  a summary of hemisphere-integrated excess (PbPb--pp) yields from all assessed sources for balanced and unbalanced dijet events is shown in Figs.~\ref{fig:Integral_Summary} and~\ref{fig:Integral_Summary_Peripheral}.  The top panels of Fig.~\ref{fig:Integral_Summary} present total central PbPb--pp differences in \pttrk associated with the subleading (plotted positive) and leading (plotted negative) jets.  The leading and subleading jet peak modifications offset each other, so the  total jet-peak-related modification, constructed from these two distributions, is also presented.  The total jet peak modifications in central PbPb collisions are not significantly different in unbalanced versus balanced dijet events.  The bottom panels of Fig.~\ref{fig:Integral_Summary} present these jet-peak modifications together with the long-range modifications evident in Fig.~\ref{fig:Integral_Longrange} to show the decomposed hemisphere-wide differences in associated \pt in each track \pt range.  Unlike the jet peak contributions, the long-range PbPb versus pp modifications are greater than their uncertainties between balanced and unbalanced dijet events.  Here the depletion of high-\pt tracks in unbalanced PbPb versus pp dijet events corresponds to the reduced contribution from additional jets (which are prominently evident in the long-range distribution for pp unbalanced dijet events) in central PbPb unbalanced dijet events.  Figure~\ref{fig:Integral_Summary_Peripheral} presents the same hemisphere-integrated PbPb--pp excess for peripheral collisions for comparison to the central results shown in Fig.~\ref{fig:Integral_Summary}.  Some possible small modifications are already evident in this 50--100\% centrality range, but these differences between peripheral PbPb and pp results are in most cases smaller than systematic uncertainties.

\clearpage

\clearpage
\section{Summary}
\label{sec:Summary}

In this analysis, the redistribution of momentum in dijet events is studied via two-dimensional $\Delta\eta$--$\Delta\phi$ jet-track correlations in PbPb and pp collisions at $\sqrt{s_\mathrm{NN}} = 2.76$\TeV, using data sets with integrated luminosities of 166\mubinv and 5.3\pbinv, respectively.  Events are selected to include a leading jet with $p_\mathrm{T,1} > 120$\GeV and subleading jet with $p_\mathrm{T,2}>50$\GeV, with an azimuthal separation of at least $\Delta\phi_{1,2} > 5\pi/6$.  Subtracting the long-range part of the correlation from the jet peaks, this work extends the studies of jet shape modifications in PbPb events relative to pp collisions to large radial distances ($\Delta r \approx 1$) from the jet axis.  These modifications are found to extend out to the largest radial distance studied in central PbPb events for both leading and subleading jets.  The jet modifications are further studied differentially for balanced and unbalanced dijet events, and as a function of \pttrk and collision centrality for each of these selections in PbPb and pp reference data.  Transverse momentum redistribution around the subleading and leading jets, as well as differences in long-range correlated background asymmetry, are separately analyzed.

An excess of transverse momentum carried by soft particles ($\pttrk < 2$\GeV) is found for both leading and subleading jets in central PbPb collisions relative to the pp reference, consistent with previous studies of charged-particle yields correlated to high-\pt jets.   For unbalanced dijet events, this low-\pttrk excess is greater on the subleading-jet side than on the leading-jet side.  However, the difference in \pttrk contained in the Gaussian-like jet peaks is found only partially to account for the total \pttrk redistribution in the most central PbPb collisions with dijet events.  In the long-range correlated distribution of \pttrk under the jet peaks, it is found that the excess of relatively high-\pt particles ($2 < \pttrk < 8$\GeV) on the subleading side relative to the leading side, which is observed in asymmetric pp collisions and mainly attributed to three-jet events, is absent in most central PbPb collisions.  This indicates that the fraction of events with additional jets in the asymmetric dijet sample is significantly lower than in an identical selection of dijet events in pp data. A long-range asymmetry in the low-\pt particles in dijet events is also observed, providing further input for the theoretical understanding of jet-medium coupling.

\clearpage
\begin{acknowledgments}
We congratulate our colleagues in the CERN accelerator departments for the excellent performance of the LHC and thank the technical and administrative staffs at CERN and at other CMS institutes for their contributions to the success of the CMS effort. In addition, we gratefully acknowledge the computing centres and personnel of the Worldwide LHC Computing Grid for delivering so effectively the computing infrastructure essential to our analyses. Finally, we acknowledge the enduring support for the construction and operation of the LHC and the CMS detector provided by the following funding agencies: BMWFW and FWF (Austria); FNRS and FWO (Belgium); CNPq, CAPES, FAPERJ, and FAPESP (Brazil); MES (Bulgaria); CERN; CAS, MoST, and NSFC (China); COLCIENCIAS (Colombia); MSES and CSF (Croatia); RPF (Cyprus); SENESCYT (Ecuador); MoER, ERC IUT and ERDF (Estonia); Academy of Finland, MEC, and HIP (Finland); CEA and CNRS/IN2P3 (France); BMBF, DFG, and HGF (Germany); GSRT (Greece); OTKA and NIH (Hungary); DAE and DST (India); IPM (Iran); SFI (Ireland); INFN (Italy); MSIP and NRF (Republic of Korea); LAS (Lithuania); MOE and UM (Malaysia); BUAP, CINVESTAV, CONACYT, LNS, SEP, and UASLP-FAI (Mexico); MBIE (New Zealand); PAEC (Pakistan); MSHE and NSC (Poland); FCT (Portugal); JINR (Dubna); MON, RosAtom, RAS and RFBR (Russia); MESTD (Serbia); SEIDI and CPAN (Spain); Swiss Funding Agencies (Switzerland); MST (Taipei); ThEPCenter, IPST, STAR and NSTDA (Thailand); TUBITAK and TAEK (Turkey); NASU and SFFR (Ukraine); STFC (United Kingdom); DOE and NSF (USA).

IIndividuals have received support from the Marie-Curie programme and the European Research Council and EPLANET (European Union); the Leventis Foundation; the A. P. Sloan Foundation; the Alexander von Humboldt Foundation; the Belgian Federal Science Policy Office; the Fonds pour la Formation \`a la Recherche dans l'Industrie et dans l'Agriculture (FRIA-Belgium); the Agentschap voor Innovatie door Wetenschap en Technologie (IWT-Belgium); the Ministry of Education, Youth and Sports (MEYS) of the Czech Republic; the Council of Science and Industrial Research, India; the HOMING PLUS programme of the Foundation for Polish Science, cofinanced from European Union, Regional Development Fund, the Mobility Plus programme of the Ministry of Science and Higher Education, the National Science Center (Poland), contracts Harmonia 2014/14/M/ST2/00428, Opus 2013/11/B/ST2/04202, 2014/13/B/ST2/02543 and 2014/15/B/ST2/03998, Sonata-bis 2012/07/E/ST2/01406; the Thalis and Aristeia programmes cofinanced by EU-ESF and the Greek NSRF; the National Priorities Research Program by Qatar National Research Fund; the Programa Clar\'in-COFUND del Principado de Asturias; the Rachadapisek Sompot Fund for Postdoctoral Fellowship, Chulalongkorn University and the Chulalongkorn Academic into Its 2nd Century Project Advancement Project (Thailand); and the Welch Foundation, contract C-1845.
\end{acknowledgments}
\bibliography{auto_generated}\cleardoublepage \appendix\section{The CMS Collaboration \label{app:collab}}\begin{sloppypar}\hyphenpenalty=5000\widowpenalty=500\clubpenalty=5000\textbf{Yerevan Physics Institute,  Yerevan,  Armenia}\\*[0pt]
V.~Khachatryan, A.M.~Sirunyan, A.~Tumasyan
\vskip\cmsinstskip
\textbf{Institut f\"{u}r Hochenergiephysik der OeAW,  Wien,  Austria}\\*[0pt]
W.~Adam, E.~Asilar, T.~Bergauer, J.~Brandstetter, E.~Brondolin, M.~Dragicevic, J.~Er\"{o}, M.~Flechl, M.~Friedl, R.~Fr\"{u}hwirth\cmsAuthorMark{1}, V.M.~Ghete, C.~Hartl, N.~H\"{o}rmann, J.~Hrubec, M.~Jeitler\cmsAuthorMark{1}, A.~K\"{o}nig, I.~Kr\"{a}tschmer, D.~Liko, T.~Matsushita, I.~Mikulec, D.~Rabady, N.~Rad, B.~Rahbaran, H.~Rohringer, J.~Schieck\cmsAuthorMark{1}, J.~Strauss, W.~Treberer-Treberspurg, W.~Waltenberger, C.-E.~Wulz\cmsAuthorMark{1}
% ...
\vskip\cmsinstskip
% ...
1:~~Also at Vienna University of Technology, Vienna, Austria\\
% ...

\end{sloppypar}
\end{document}